\documentclass[]{svmult1}

\usepackage{type1cm}        
\usepackage{makeidx}         
\usepackage{graphicx}        
\usepackage{multicol}        
\usepackage[bottom]{footmisc}

\usepackage{cite}
\usepackage{epsfig}
\usepackage{graphicx}
\usepackage{amsmath}
\usepackage{amssymb}
\usepackage{ulem}
\usepackage{mdwlist}
\usepackage{color}

\usepackage{color}
\usepackage{slashed}

\makeindex             

\newcommand{\Mvec}{{\rm\bf M}}
\newcommand{\ep}{\varepsilon}
\newcommand{\Li}{{\rm Li}}
\newcommand{\HA}{{\rm H}}
\newcommand{\shuffle}{\, \raisebox{1.2ex}[0mm][0mm]{\rotatebox{270}{$\exists$}} \,}

\allowdisplaybreaks[4]

\begin{document}

\title*{{\footnotesize{\sf DESY 21--038,~~DO-TH 21/09,~~SAGEX-21-06
}}\\
Analytic integration methods in quantum \\ field theory: an Introduction
\protect\footnote{Contribution to the 
Volume ``Antidifferentiation and the Calculation of Feynman Amplitudes'', Springer, Berlin, 2021.
}}
\titlerunning{Analytic Integration Methods in QFT} 
\author{Johannes Bl\"umlein}
\institute{Deutsches Elektronen-Synchrotron, DESY, Platanenallee 6, D-15738 Zeuthen, Germany, 
\email{Johannes.Bluemlein@desy.de}}
%
%
\maketitle

\abstract{
A survey is given on the present status of analytic calculation methods and the mathematical 
structures of zero- and single scale Feynman amplitudes which emerge in higher order perturbative 
calculations in the Standard Model of elementary particles, its extensions and associated model 
field theories, including effective field theories of different kind.}

\section{Introduction}
\label{sec:1}

\vspace{1mm}\noindent
Analytic precision calculations for the observables in renormalizable quantum field theories have 
developed during the last 70 years significantly. These methods have helped to put the Standard
Model of elementary particles to tests of an unprecedented accuracy, requested by the scientific
method \cite{VELTMAN}. Present and future high luminosity experiments \cite{EXP,Abada:2019zxq} will 
demand 
even higher precision predictions at the theory side. This goes along with mastering large sets 
of analytic data by methods of computer algebra and special mathematical methods to perform the
corresponding integrals analytically.

Here analytic integration is understood as antidifferentiation. In this context the first question
going to arise is: which is the space to represent a certain class of integrals in an irreducible 
manner. As history showed, this question is usually answered in an iterative way. Often not all
the existing relations in a given class of functions can be revealed right from the beginning. 
It is even so that in some cases it remained unclear over many decades whether all 
relations are already found or not. One example for this are the multiple zeta values \cite{Blumlein:2009cf}.
The development of physics applications is of course not stopped by this, as also partial solutions
are of great help in reducing the large complexity to be dealt with. The application of quite different 
techniques has often to be combined to finally tackle these cutting edge problems. Moreover, it is this
process which delivers new insights and is able to produced even more refined technologies.

The problem of analytic integration of Feynman diagrams is nowadays a field of research requesting 
to join fundamental ideas from theoretical physics, computer algebra and computing technology, as well as
of a growing number of fields in pure mathematics. The topics are so challenging that the experts in all these
different fields are attracted by them in the solution of the different problems.

The present interdisciplinary workshop, organized by the Wolfgang-Pauli Center, arose from the idea to get
scientists working in the field of Quantum Field Theory, computer algebra and pure mathematics
together to review the status of the analytic solution of Feynman integrals reached and to prepare for 
further developments.

The topics of the workshop included
both techniques to reduce the number of Feynman diagrams by physical relations, such as the integration 
by parts relations \cite{MARQUARD,VERMASEREN, FRELLESVIG} as well as the mathematical methods to compute 
these integrals analytically. The latter include the method of generalized hypergeometric functions 
\cite{KALMYKOV} and the general theory of contiguous relations \cite{PAULE}, the methods of integer relations 
\cite{BROADHURST}, guessing methods of one--dimensional quantities, hyperlogarithms \cite{PANZER}, the solution of 
master-integrals using difference and differential equations \cite{SCHNEIDER,WEIL,KOTIKOV,HENN}, Risch algorithms 
on nested integrals and rationalization algorithms \cite{RAAB}, holonomic integration \cite{KOUTSCHAN}, the 
multivalued Almkvist-Zeilberger algorithm \cite{ABLINGER}, expansion by regions \cite{SMIRNOV}, elliptic integrals 
and related topics \cite{WEINZIERL,BROEDEL}, cutting techniques \cite{KREIMER}, and special multi-leg applications 
\cite{HIPPEL,BARTELS,Papathanasiou}. In different precision calculations these technologies are applied.\footnote{
For a summary on recent massless calculations, see \cite{MOCH}.}

In this paper we give a brief introduction into the topic\footnote{For other recent surveys on integration methods 
for Feynman integrals see \cite{Weinzierl:2010ps,Ablinger:2013eba,Ablinger:2013jta,Weinzierl:2013yn,Duhr:2014woa,
Blumlein:2018cms}.}, covering the main steps in multi-loop perturbative calculations in 
Section~\ref{sec:2}. Then we turn to the different symbolic integration techniques of Feynman parameter integrals in
Section~\ref{sec:3} and describe the associated function spaces in Section~\ref{sec:4}. All these technologies serve
the purpose to reach a higher theoretical precision for many observables in Quantum Field Theory to cope with the 
experimental precision data and to either confirm the Standard Model of elementary particles to higher accuracy
or to find signals of new physics. Some aspects of this are discussed in Section~\ref{sec:5} and Section~\ref{sec:6} 
contains the conclusions.
\section{Principle computation steps for Feynman diagrams}
\label{sec:2}

\vspace{1mm}\noindent
In any large scale calculation there is the need to generate the Feynman diagrams in an automated 
way. One of the important packages to provide this is {\tt QGRAF} \cite{Nogueira:1991ex}, for which
the corresponding physics file containing the Feynman rules has to be provided. Necessary group-theoretic
calculations, as e.g. the color algebra in QCD, can be carried out using the package {\tt Color}
\cite{vanRitbergen:1998pn}. Finally, there is the necessity to perform the Dirac- and Lorentz-algebra,
in an efficient manner, which is provided by {\tt Form} \cite{FORM}. 

Next, the integration by parts reduction \cite{IBP} has to be performed. For many processes up to three 
loops in QCD the Laporta algorithm \cite{Laporta:2001dd}  in its different implementations 
\cite{Smirnov:2008iw,Smirnov:2019qkx,Studerus:2009ye,vonManteuffel:2012np,MARSEID,Maierhoefer:2017hyi,
MARQUARD} is sufficient. At even higher orders the complexity becomes larger and larger
and it is necessary to combine different methods or to device algorithms tailored to the particular problem 
to be solved \cite{VERMASEREN,MARQUARD,FRELLESVIG}. There will be certainly more developments in this
field in the future.

After this reduction one obtains the master integrals, which represent the quantity to be finally 
calculated, and it has to be decided in which way the computation shall be put forward. One way,
if it can be pursued, is to calculate the individual master integrals to the necessary depth in the dimensional 
parameter $\ep = D-4$, with $D$ the dimension of the space-time using the analytic calculation 
technologies described in Section~\ref{sec:3}. At lower orders in the coupling constant up to moderate
complexity in the involved mass scales this is possible. It may even be that a single technology,
like the calculation of the master integrals by solving the associated differential equations provides
the full solution.

However, in particular in massive calculations it is possible, that the master integrals contain elliptic 
parts at 3--loop order, but the quantity to be calculated is known to be free of these contributions.
In such a case one may use the method of arbitrarily high Mellin moments for single scale quantities
\cite{Blumlein:2017dxp} to express the moments of the master integrals. Here the elliptic structures
are fully encoded in just rational coefficients. One forms the observable to be calculated which are
given as series of Mellin moments weighted by $\zeta$- and color factors and here the elliptic or hyperelliptic
contributions cancel. The method of guessing \cite{Blumlein:2009tj,GSAGE} will then enable one 
to find the recurrence relation for the complete result. This method requires a large number of moments, which, 
however, can be algorithmically
provided \cite{Blumlein:2017dxp}. Recent applications are \cite{Ablinger:2017tan,Ablinger:2017ptf,Behring:2019tus,
Blumlein:2019oas}.
The solution of the recurrences provided by guessing is using difference ring theory \cite{DRING} implemented in the
package {\tt Sigma}  \cite{Schneider:2007a,Schneider:2013a} in the case that the recurrences factorize to first 
order. Other cases are discussed in Sections~\ref{sec:3} and \ref{sec:4}.

Let us now turn to the specific antidifferentiation methods for Feynman integrals.
\section{Symbolic Integration of Feynman Parameter Integrals}
\label{sec:3}

\vspace{1mm}\noindent
Most of the analytic methods described in this section have more general applications than just 
to be used for the evaluation of Feynman integrals and were developed even without knowing
of this particular application. Still the challenges to integrate also involved Feynman diagrams
have refined many of these methods significantly. Non of these methods is universal and it is
often an appropriate combination of these methods leading to optimal solutions of a project 
at hand, w.r.t. the necessary computational requests such as memory, storage and computational
time in the case of the high end calculations.

Many of the problems can be cast into discrete formulations, allowing to make use of methods 
of difference ring theory. Here the packages {\tt Sigma}
\cite{Schneider:2007a,Schneider:2013a}, {\tt EvaluateMultiSums} and {\tt SumProduction}
\cite{Ablinger:2010pb,Blumlein:2012hg,Schneider:2013zna}, see also \cite{Schneider:19},
can be used.

In the following we describe the PSLQ method for zero-dimensional quantities, hypergeometric functions
and their generalizations, analytic solutions using Mellin--Barnes integrals, hyperlogarithms, guessing techniques,
the method of difference and differential equations, and the Almkvist-Zeilberger algorithm.
\subsection{PSLQ: zero-dimensional integrals}
\label{sec:23}

\vspace{1mm}\noindent
In expanding perturbatively in the coupling constant several physical quantities are zero-dimensional,
i.e. they can be represented by numbers only. A recent example consists in the QCD $\beta$-function, now known 
to 5--loop order \cite{BETA}. The respective expressions are given by the 
color factors of the  gauge group, rational terms and special numbers, as e.g. multiple zeta values 
\cite{Borwein:1999js,Blumlein:2009cf}. If one knows the potential pool of all the contributing special numbers 
one may try to determine the rational coefficients of the whole problem by providing enough numerical digits 
for the corresponding quantity. One method to obtain such an experimental result is PSLQ \cite{PSLQ}.

One example is given by determining the integral
\begin{eqnarray}
I_1 = \int_0^1 dx \frac{\Li_2(x)}{1+x} \approx 0.3888958461681063290997435080476931009885,
\end{eqnarray}
where $\Li_2(x)$ denotes the classical dilogarithm \cite{POLYLOG}. $I_1$ is
a weight {\sf w = 3} multiple zeta value \cite{Blumlein:2009cf} for which the basis is known. It is spanned by  
\begin{eqnarray}
\{\ln^3(2), \zeta_2 \ln(2), \zeta_3\}.
\end{eqnarray}
PSLQ delivers the following representation
\begin{eqnarray}
I_1 = \ln^3(2) - \frac{5}{8} \zeta_3,
\end{eqnarray}
which is also obtained by a direct analytic calculation. In any case it is important to have enough 
digits available. If a result has been obtained it should be verified by an even larger number of 
digits. The results usually remain experimental. In many complex applications it is difficult 
to proof the result analytically. Advanced applications of these and similar methods are discussed in 
\cite{BROADHURST}.
\subsection{Generalized hypergeometric functions and their extensions}
\label{sec:31}

\vspace{1mm}\noindent
The integrands of multi--dimensional Feynman parameter integrals are hyperexponential, i.e. given by products 
of multivariate polynomial expressions raised to real powers, implied by the dimensional parameter $\ep$. These type of 
functions correspond to the integrands
defining the (generalized) hypergeometric functions \cite{HYPKLEIN,HYPBAILEY,SLATER1} and their generalizations such as 
the Appell-, Kampe-De-Feriet- and related functions \cite{APPEL1,APPEL2,KAMPE1,EXTON1,EXTON2,SCHLOSSER,Anastasiou:1999ui,
Anastasiou:1999cx,SRIKARL,Lauricella:1893,Saran:1954,Saran:1955}. The advantage of these integral representations is
that they usually have a lower dimensional series representation compared to their integral representations and
a part of the original integrals can be performed in this way. The simplest function is Euler's Beta-function implying 
the series of $_{p+1}F_p$ functions
\begin{eqnarray} 
B(a_1,a_2)           &=& \int_0^1 dt~t^{a_1-1} (1-t)^{a_2-1}
\\
_3F_2(a_1,a_2,a_3;b_1,b_2;x) &=& \frac{\Gamma(b_2)}{\Gamma(a_3) \Gamma(b_2-a_3)} \int_0^1 dt~t^{a_3-1}
(1-t)^{-a_3+b_2-1}
\nonumber\\ && \times
_2F_1(a_1,a_2;b_1;tx).
\end{eqnarray}
Up to the level of the massless and massive two--loop calculations for single--scale quantities in QCD these 
representations are usually sufficient \cite{Hamberg:1990np,HAMBERG,Buza:1995ie,Bierenbaum:2007qe}. 
In the case of three--loop ladder graphs also Appell-functions \cite{Ablinger:2012qm,Ablinger:2015tua} 
contribute. 
A survey on the status of this method has been given in \cite{KALMYKOV}. In relating the different special 
functions of this kind contiguous relations play an essential role, which has been discussed in \cite{PAULE} in 
detail.
One ends up with a series of infinite sum representations, for which the $\ep$--expansion is performed.
These sums have to be further dealt with by using summation methods, cf.~Section~\ref{sec:35}. 
\subsection{The analytic Mellin--Barnes technique}
\label{sec:32}

\vspace{1mm}\noindent
Only the simpler hyperexponential integrands can be represented by the higher transcendental functions
described in Section~\ref{sec:31}. One major problem to proceed are the structures of some of the 
hyperexponential 
factors, for which the contributing variables cannot be cast into a form required in the previous case.
Here the use of Mellin--Barnes integrals \cite{BARNES1,MELLIN1} is of help, which are
defined by
\begin{eqnarray} 
\frac{1}{(a+b)^\alpha} = \frac{1}{\Gamma(\alpha)} \frac{1}{2\pi i} \int_{-i \infty}^{i \infty} dz \Gamma(\alpha + z)
\Gamma(-z) \frac{b^z}{a^{\alpha+z}}, ~~~\alpha \in \mathbb{R}, \alpha > 0,
\end{eqnarray} 
cf. e.g.~\cite{SMIRNOV1}. Here the contour integral is understood to be either closed to the left or the right
surrounding the corresponding singularities. Note that also the functions in Section~\ref{sec:31} have 
representations in terms of Pochhammer--Umlauf integrals \cite{POCHHAMMER,HYPKLEIN,KF} and therefore Mellin--Barnes
integral representations.
The different Mellin--Barnes integrals can be turned into a number of infinite series by the residue theorem,
leading to nested sums to be dealt with further by the summation technologies implemented in the package 
{\tt Sigma}~\cite{Schneider:2007a,Schneider:2013a}. Here it is not a priori clear that all the sums can be solved,
which will turn out by working through the algorithm. One is advised therefore not to use the Mellin--Barnes
splitting of the integrands extensively, although being possible \cite{Blumlein:2010zv}. On the other hand, one may apply 
the packages 
for Mellin--Barnes integrals \cite{Czakon:2005rk,Smirnov:2009up,Gluza:2007rt,Gluza:2010rn} to obtain numerical results
for comparisons, to check the final analytic results. One reason that a summation problem cannot be solved completely
is related to the fact, that the associated recurrences are not first order factorizing and other technologies have to 
be applied. Mellin--Barnes integrals do significantly extend the methods described in Section~\ref{sec:31} and may lead
to new higher transcendental functions not known yet in the literature.
\subsection{Hyperlogarithms}
\label{sec:33}

\vspace{1mm}\noindent
The idea behind the method of hyperlogarithms is that for certain multivariate Feynman parameter integrals 
an order of integrations can be found in which the respective parameters to be integrated over always occur in 
linear form (Fubini sequence) \cite{Brown:2008um}. In this way the corresponding integrals are cast into 
iterative Kummer--Poincar\'e integrals \cite{KUMMER,POINCARE,LADAN,CHEN,GONCHAROV}. Originally the method 
could only be applied to non--singular integrals in the dimensional parameter $\ep$ and an extension has been 
worked out in \cite{vonManteuffel:2014qoa} to the singular case. An implementation of the algorithm has been 
given in \cite{Panzer:2014caa}. The method
has first been applied to massless Feynman integrals. A generalization for massive integrals,
also containing local operator insertions, has been given in \cite{Ablinger:2014yaa,Wissbrock:2015faa},
where multi-linearity is broken in part, still yielding analytic results. The method is interesting but of limited use, 
since 
it applies to structurally 
simple cases only and requires more than just first order factorization of the associated differential equations
or the related nested sum representations, through the application of which much more general cases can be 
solved. 
\subsection{The method of guessing}
\label{sec:34}

\vspace{1mm}\noindent
The integral transform
\begin{eqnarray} 
\Mvec[f(x)](N) = \int_0^1 dx x^{N-1} f(x)
\end{eqnarray} 
defines the Mellin transform, which will often appear in the following. If singlevariate multiple Feynman parameter 
integrals $f(x)$ in QCD processes are viewed in terms of their 
Mellin moments $\{\left. a(N)\right|_{N=1}^\infty\}$ for fixed values of $N$, one obtains series of rational 
numbers 
weighted by color factors and multiple zeta values or other special numbers, cf.~\cite{Larin:1993vu,Larin:1996wd,
Retey:2000nq,Blumlein:2004xt}. It turns out in very many practical cases that the general $N$ solution generating 
the individual moments $a(N)$ obey recursion relations. This is the case e.g. for (massive) operator matrix 
elements \cite{Bierenbaum:2009mv} but also for single--scale Wilson coefficients, Ref.~\cite{Vermaseren:2005qc}.

One would like now to determine this recurrence on the basis of a (large) number of these moments 
algorithmically. The corresponding algorithms are called guessing methods \cite{GSAGE}, which are also available 
in {\tt Sage} \cite{SAGE}, exploiting the fast integer algorithms available there. The method returns the 
wanted difference equation, and tests it by a larger series of 
further moments. This method has been applied in Ref.~\cite{Blumlein:2009tj} to obtain from more than 5000 
moments the massless unpolarized three--loop anomalous dimensions and Wilson coefficients in deep-inelastic 
scattering \cite{ANOMDWIL,Vermaseren:2005qc}. More recently, the method has been applied {\it ab 
initio} in the calculation of three--loop splitting functions \cite{Ablinger:2017tan,Behring:2019tus} and the massive 
two-- and 
three--loop form factor \cite{Ablinger:2018zwz,Ablinger:2018yae}. The largest systems solved in this context 
were massive operator matrix elements needing $\sim 8000$ moments \cite{Ablinger:2017ptf} 
to derive the corresponding recurrences.

One then tries to solve these recurrences with the package {\tt Sigma} \cite{Schneider:2007a,Schneider:2013a},
which will either find the solution or does at least factor off all the first order factors, separating the
remaining part to be solved using other techniques. The large number of moments needed is generated using
the method described in Ref.~\cite{Blumlein:2017dxp}. This algorithm will play a central role in many
upcoming calculations in the singlevariate case.

Other algorithms such as {\tt Mincer} \cite{Gorishnii:1989gt}, {\tt MATAD} \cite{Steinhauser:2000ry}
or {\tt Q2E} \cite{Harlander:1997zb,Seidensticker:1999bb} do also provide Mellin moments. However, the number
of moments which can be obtained with these formalisms is rather low. Still these packages play a very 
essential role in higher order calculations, since they provide independent tests and they are used both for 
predicting intermediary and final results for indispensable comparisons.
\subsection{Difference equations and summation methods} 
\label{sec:35} 

\vspace{1mm}\noindent
Many of the problems occurring in analytic Feynman integral calculation can be mapped to summation 
problems and the solution of difference equations. Infinite and finite sums appear in binomial and 
Mellin--Barnes decompositions and also in the expansion of Pochhammer symbols depending on the dimensional
parameter $\ep$ into the associated Laurent series. Moreover, ordinary differential equations in 
a variable $x$ can be transformed into recursions in the variable $N$ by a Mellin transform \cite{NOERLUND}
Furthermore, structures in $x$--space can be expanded into formal Taylor series, the $N$th coefficient 
of which, $a(N)$, also obeys a certain recurrence. 

Nested sums over hypergeometric terms $h(k)$, with $h(k+1)/h(k) \in \mathbb{Q}(k)$, will form the basis of 
the summation problems we briefly consider in the following. One first considers finite sums, i.e. those terminating
at an upper integer for all summation quantifiers. The corresponding sums are then cast into the form
\begin{eqnarray} 
\label{eq:SUM}
S_{b,\vec{a}}(N;\vec{c}) = \sum_{k=1}^N h_b(k;\vec{c}) S_{\vec{a}}(k;\vec{c}).
\end{eqnarray} 
Here $\{\vec{c}\}$ denotes a finite set of constants, which has to be added to the ground field of the difference 
ring.
These are also given by certain physical parameters in multi--scale processes.
Infinite sums can be dealt with by considering limiting procedures implemented in the package {\tt HarmonicSums}
\cite{HARMSU,
Blumlein:2009ta,Ablinger:2013hcp,Vermaseren:1998uu,Blumlein:1998if,Remiddi:1999ew,Ablinger:2011te,Ablinger:2013cf,Ablinger:2014bra}.
In solving a summation problem at hand the associated sum-algebra is built and the corresponding sums appearing in the
final result are simplified accordingly.

All summation problems which lead to first order factorizing recurrences can be solved using 
difference ring theory \cite{DRING} implemented in the package {\tt Sigma} \cite{Schneider:2007a,Schneider:2013a}.
This concerns a rather wide class of cases. By the systematic use of these techniques, harmonic sums 
\cite{Vermaseren:1998uu,Blumlein:1998if} generalized harmonic sums \cite{Ablinger:2013cf}, cyclotomic harmonic sums 
\cite{Ablinger:2011te}, and finite binomial and inverse--binomial sums \cite{Ablinger:2014bra}
can be dealt with. A part of these function spaces has been found and systematically explored by these techniques.
Recent developments in this field are summarized in Ref.~\cite{SCHNEIDER}.

These methods also apply to cases, which are not factorizing to first order, as e.g. in Ref.~\cite{Ablinger:2017ptf}.
Here all first order factors are separated from a remainder non--factorizing recurrence. The latter one
can be further dealt with using different techniques.

\vspace*{-0.7cm}
\subsection{Differential equations}
\label{sec:36}

\vspace{1mm}\noindent
The IBP-relations \cite{IBP} do naturally imply systems of differential equations for the master integrals.
In the case of single--scale quantities these are systems of ordinary differential equations, Eq.~(\ref{eq:DEQ1}),
which have to be solved, providing the necessary boundary conditions. Early investigations following this approach were
\cite{Kotikov:1990kg,Bern:1992em,Remiddi:1997ny,Gehrmann:1999as}. 
\begin{eqnarray}
\label{eq:DEQ1}
\frac{d}{dx}
\left(
\begin{array}{c}
f_1\\ \vdots \\ f_n\end{array}\right)
= \left(\begin{array}{ccc}
A_{11} & \hdots & A_{1,n}
\\
\vdots &  & \vdots
\\
A_{n1} & \hdots & A_{n,n}
\end{array} \right)
\left(\begin{array}{c}f_1\\ \vdots \\ f_n\end{array}\right)
+ \left(\begin{array}{c}g_1\\ \vdots \\ g_n\end{array}\right),
\end{eqnarray}

One way to solve the system (\ref{eq:DEQ1}) consists in decoupling it using the methods of 
\cite{BCP13,Zuercher:94} encoded in {\tt Oresys} \cite{ORESYS}. In this way one obtains one scalar 
differential equation of higher order 
\begin{eqnarray}
\label{eq:DEQ2}
\sum_{k=0}^n p_{n-k}(x)\frac{d^{n-k}}{dx^{n-k}} f_1(x) = \overline{g}(x),
\end{eqnarray}
with $p_n\neq0$, and $(n-1)$ equations for the remaining solutions, which are fully determined by the solution $f_1(x)$.
One also may transform Eq.~(\ref{eq:DEQ1}) into Mellin space, decouple there and solve using the  
efficient methods of the package {\tt Sigma}, cf.~Ref.~\cite{Ablinger:2015tua}.

In the case of first order factorization
the decoupled differential operator of (\ref{eq:DEQ2}) can be written in form of a combination of iterative 
integrals, cf.~Section~\ref{sec:42},
\begin{eqnarray}
\label{eq:DEQ4}
f_1(x) &=& \sum_{k = 1}^{n+1} \gamma_k g_{k}(x),~\gamma_k \in {\mathbb C},\\
g_{k}(x) &=& h_0(x)\int_0^x dy_1 h_{1}(y_1) \int_0^{y_1} dy_2 h_{2}(y_2) ...
\int_0^{y_{k-2}} dy_{k-1} h_{k-1}(y_{k-1})
\nonumber\\ && \times
\int_0^{y_{k-1}} dy_{k} q_k(y_{k}),
\end{eqnarray}
with $q_k(x)=0$ for $1\leq k\leq m$. Further, $\gamma_{m+1}=0$ if $\bar{g}(x)=0$ in~\eqref{eq:DEQ2}, and
$\gamma_{m+1}=1$ and $q_{m+1}(x)$ being a mild variation of $\bar{g}(x)$ if $\bar{g}(x)\neq0$. 

One obtains d'Alembertian solutions~\cite{Abramov:94} since the master integrals appearing in quantum field theories 
obey 
differential equations with rational coefficients, the letters $h_i$, which constitute the iterative integrals, 
have to be hyperexponential and the solution can be computed using the package~\texttt{HarmonicSums}. 
Liouvillian solutions~\cite{Singer:81} can also be calculated with~\texttt{HarmonicSums} utilizing Kovacic's 
algorithm~\cite{Kovacic:86}. This algorithm has been applied in various massive three--loop calculations so far, 
cf.~\cite{Ablinger:2015tua,Ablinger:2017ptf,Ablinger:2017hst}.
A solution algorithm for first order systems has also been presented in \cite{Ablinger:2018zwz}.
In these algorithms no specific choice of a basis is necessary and the different contributions in the 
$\ep$-expansion are obtained straightforwardly.

In the multivariate case, the so-called $\varepsilon$--representation of a linear system of partial 
differential equations
\begin{eqnarray}
\label{eq:DEQ5a}
\partial_m f(\varepsilon, x_n) = A_m(\varepsilon, x_n) f(\varepsilon, x_n)
\end{eqnarray}
is important, as has been recognized in  Refs.~\cite{Kotikov:2010gf,Henn:2013pwa}, see also \cite{Henn:2014qga}.
The matrices $A_n$ can now be transformed in the non--Abelian case by
\begin{eqnarray}
A_m' = B^{-1} A_m B - B^{-1}(\partial_m B),
\end{eqnarray}
as well--known \cite{NOVIKOV:80,Sakovich:1995}. One then intends to find a matrix $B$ to transform 
(\ref{eq:DEQ5a}) into the form
\begin{eqnarray}
\label{eq:DEQ6}
\partial_m f(\varepsilon, x_n) = \varepsilon A_m(x_n) f(\varepsilon, x_n),
\end{eqnarray}
if possible. This then yields solutions in terms of iterative integrals. Here a formalism for the basis change 
to the $\varepsilon$--basis has been proposed in \cite{Lee:2014ioa} and implemented in the singlevariate case 
in \cite{Prausa:2017ltv,Gituliar:2017vzm} and in the multivariate case in \cite{Meyer:2017joq}. 
All these methods apply only if the systems are first order factorizable.

In the solution of differential equations it is often important to rationalize roots as much as possible 
\cite{Besier:2018jen,Besier:2019kco,RAAB}. The corresponding algorithms are important for the derivation of 
iterated root--valued 
integrals also in the multivariate case \cite{ISR,ROOT1}. The corresponding structures do then allow expansions 
in small parameters to obtain even more compact analytic results, since normally the iterated integrals
with various involved root--valued letters turn out to form very large expressions. Recent developments have been 
summarized in \cite{RAAB}. General aspects on the solution of differential equation systems were summarized in 
Ref.~\cite{KOTIKOV}, while aspects of differential Galois theory were discussed in \cite{WEIL}.
In the solution
of ordinary differential equations emerging in the context of Feynman diagrams holonomic integration
often provides a powerful tool, see.~\cite{KOUTSCHAN}.
\subsection{Multivalued Almkvist-Zeilberger algorithm}
\label{sec:37}

\vspace{1mm}\noindent
Singlevariate Feynman parameter integrals $I(x,\ep)$ are integrals over $\{x_i|_{1=1}^n\}$ $\in [0,1]^n$
with one more free parameter $x \in [0,1]$ and the dimensional parameter $\ep$. A Mellin transform 
leads to the function $\hat{I}(N,\ep)$. The Almkvist--Zeilberger algorithm \cite{AZ1,AZ2} provides a method to 
either find an associated differential equation for $I(x)$ or a difference equation for $I(N)$, the 
coefficients of which are either polynomials in $\{x,\ep\}$ or $\{N,\ep\}$, 
\begin{eqnarray}
\sum_{l=0}^m P_l(x,\varepsilon) \frac{d^l}{dx^l} I(x,\varepsilon) &=& N(x,\varepsilon)
\label{eq:AZ1}
\end{eqnarray}
\begin{eqnarray}
\sum_{l=0}^m R_l(N,\varepsilon) I(N+l,\varepsilon) &=& M(N,\varepsilon).
\label{eq:AZ2}
\end{eqnarray}

\noindent
Both equations may be inhomogeneous, where the inhomogeneities emerge as known functions from lower order problems.
An optimized and improved algorithm for the input class of Feynman integrals has
been implemented in the \texttt{MultiIntegrate} package~\cite{Ablinger:2013hcp,Ablinger:2015tua}.
It can either produce homogeneous equations of the form~(\ref{eq:AZ1},\ref{eq:AZ2}) or equations with an 
inhomogeneity formed out of already known functions. The method extends successively the structural form of the 
difference or differential equation for the functions $\hat{I}$ or ${I}$ unless a solution is found.

This algorithm is of great use in specific cases in which either direct summation problems or the solution 
of associated differential equations becomes to voluminous or in the case of very long computation times.
Like also in the case of guessing the corresponding recurrences turn out to be well homogeneized, which makes
their solution easier.
\section{The Function Spaces}
\label{sec:4}

\vspace{1mm}\noindent
The solution of the different massless and massive higher loop calculations for zero-, single-, and 
multiscale problems induce specific function spaces, which also form algebras. These structures have been 
revealed in more detail after 1997 along with performing more and more involved computations. Before it has been known
that specific numbers play a role in Feynman integral calculations, see e.g. \cite{Broadhurst:1996kc}, 
and in the one--dimensional 
case Nielsen integrals \cite{NIELSEN}, generalizing the classical polylogarithms \cite{POLYLOG}, were in use.

The first generalizations of these functions led to nested sum structures on the one hand \cite{Vermaseren:1998uu,Blumlein:1998if}, 
and iterative integrals over certain alphabets on the other hand \cite{Remiddi:1999ew}. These structures do also apply 
to not too involved 
multi-scale problems. Iterative non-iterative integrals occurred with the advent of complete elliptic integrals
in letters, or more generally higher transcendental functions for whose integral representation the variable 
to be integrated over cannot be transformed in one of the integration boundaries only.
Synonymous objects appear in the associated sums.

In the following we describe the hierarchies of spaces for iterated integrals and nested sums,
which contribute in Feynman diagram calculations. Beyond these structures there are also
problems leading to non first order factorizable recurrences and differential operators.
In all calculations special numbers occur, which have representations by iterated integrals 
at $x=1$ or through nested sums in the limit $N \rightarrow \infty$. Even others appear in the context of
the quantities discussed in Section~\ref{sec:44}. Finally, we will also discuss numerical
representations of all these functions, including the analytic continuation of nested sums
to $N \in \mathbb{C}$.

\subsection{Nested Sums}
\label{sec:41}

\vspace{1mm}\noindent
Considering the singlevariate case sum representations have the form given in Eq.~(\ref{eq:SUM}).
Very often finite sums of another type have first to be brought into this representation using the 
algorithms encode in the package {\tt Sigma}  \cite{Schneider:2007a,Schneider:2013a}. Furthermore, also infinite sums 
have to be handled, which are usually considered as the limit $N \rightarrow \infty$ of the associated finite sums.

The sums obey quasi--shuffle relations \cite{HOFFMAN,Blumlein:2003gb}, see Section~\ref{sec:43}. The simplest 
structures are the finite
harmonic sums \cite{Vermaseren:1998uu,Blumlein:1998if}, where $g_b(k) = ({\rm sign}(b))^k/k^{|b|},~~b \in \mathbb{N} 
\backslash \{0\}$. A generalization is obtained in the cyclotomic case \cite{Ablinger:2011te}. Here the characteristic
summands are $g_{a,b,c}(k) = (\pm 1)^k/(a k +b)^{c}$, with $a,b,c \in \mathbb{N} \backslash \{0\}$.  Further, the 
generalized harmonic sums have letters
of the type $b^k/k^c$, with $c \in \mathbb{N} \backslash \{0\}, b \neq 0, b \in \mathbb{R}$, \cite{Ablinger:2013cf}.
A generalization of the last two classes of sums are those generated by the Mellin transform of iterative integrals
with letters induced by quadratic forms \cite{SQALPHA}, see Section~\ref{sec:42}.
Another generalization are nested finite binomial and inverse--binomial sums, containing also other sums discussed 
before. An example is given by
\begin{eqnarray}
\sum_{i=1}^N \frac{1}{(2i+1) \binom{2i}{i}} \sum_{j=1}^i \binom{2j}{j} \frac{(-1)^j}{j^3}
&=& \frac{1}{2} \int_0^1 \frac{(-x)^N-1}{x+1} \frac{x}{\sqrt{x + \tfrac{1}{4}}} \HA^*_{\sf w_{14},0,0}(x)
\nonumber\\
&& - \frac{ \HA^*_{\sf -\tfrac{1}{4},0,0}(0)}{2\pi} \int_0^1 dx 
\frac{\left(\tfrac{x}{4}\right)^N-1}{x-4} 
\frac{x}{\sqrt{1-x}},
\nonumber\\
\end{eqnarray}
see~\cite{Ablinger:2014bra}. Here the indices ${\sf w_k}$ label specific
letters given in \cite{Ablinger:2014bra} and the iterated integrals $\HA^*$ are defined over the support
$[x,1]$.
Infinite binomial and inverse--binomial sums have been considered in \cite{Davydychev:2003mv,Weinzierl:2004bn}.
Given the general structure of (\ref{eq:SUM}) 
many more iterated sums can be envisaged and may still appear in even higher order calculations.

\subsection{Iterated Integrals}
\label{sec:42}

\vspace{1mm}\noindent
Iterated integrals are of the form
\begin{eqnarray}
\HA_{b,\vec{a}}(x) = \int_0^x dy f_b(y) \HA_{\vec{a}}(y),~~~\HA_\emptyset = 1, f_c \in \mathfrak{A},~x 
\in[0,1],
\label{eq:ITIN}
\end{eqnarray}
where $f_c$ are real or complex--valued functions and are the letters of the alphabet $\mathfrak{A}$. 
For certain letters regularizations are required since otherwise the corresponding integrals so not exist. This 
occurs if the letters have poles in $x \in [0,1]$. Iterated integrals obey shuffle relations 
\cite{REUTENAUER,Blumlein:2003gb} which allows to represent them over a basis of fewer terms.

The simplest iterative integrals having been considered in quantum field theory are the Nielsen integrals
for the two--letter alphabets $\{1/x,1/(1-x)\}$ or $\{1/x,1/(1+x)\}$ \cite{NIELSEN}, 
covering also the polylogarithms \cite{POLYLOG}. This class has later been extended to
the harmonic polylogarithms \cite{Remiddi:1999ew} built over the alphabet $\{1/x,1/(1-x),1/(1+x)\}$. 

A further 
extension
is to the real representations of the cyclotomic polylogarithms, with $\{1/x,1/\Phi_k(x)\}$
\cite{Ablinger:2011te}, where $\Phi_k(x)$ denotes the $k$th cyclotomic polynomial. Another extension is given by
Kummer--Poincar\'e iterative integrals over the alphabet $\{1/(x-a_i),~~a_i \in \mathbb{C}\}$, 
\cite{KUMMER,POINCARE,LADAN,CHEN,GONCHAROV}. Properties of these functions have been studied in 
Refs.~\cite{Moch:2001zr,Ablinger:2013cf}. In general one may have also more general denominator polynomials $P(x)$,
which one can factor into
\begin{eqnarray}
P(x) = \prod_{k=1}^n (x - a_k) \prod_{l=1}^m (x^2 + b_l x + c_l),~~a_k, b_l, c_l \in \mathbb{R}
\end{eqnarray}
in real representations. One then performs partial fractioning for $1/P(x)$ and forms iterative integrals 
out of the obtained letters in
\begin{eqnarray}
\label{eq:AR}
\mathfrak{A}_{R} = \left\{\left. \frac{1}{x-a_i}, \frac{1}{x^2 + b_i x + c_i}, \frac{x}{x^2 + b_i x + c_i}
\right| a_i, b_i, c_i \in \mathbb{R},~ 4 c_i \geq b_i \right\},
\nonumber\\
\end{eqnarray}
cf.~\cite{SQALPHA}.

The iterated integrals (\ref{eq:ITIN}) can be analytically continued from $x \in [0,1]$ to the complex plane by 
observing their respective cuts. For the harmonic polylogarithms this has been described in 
\cite{Gehrmann:2001pz}. For the other cases the corresponding algorithm is implemented in {\tt HarmonicSums}, see also
\cite{Vollinga:2004sn}.

Further classes are found for square--root valued letters as studied e.g. in Ref.~\cite{Ablinger:2014bra}. 
In multi--scale problems, cf.~e.g. \cite{Ablinger:2017xml,ROOT1,ISR}, further 
root--valued letters appear, like also the Kummer--elliptic integrals \cite{ROOT1}, which are 
iterative integrals, because the elliptic structure is due to incomplete elliptic integrals.

The occurrence of several masses or additional external non--factorizing scales in higher order loop- 
and phase--space integrals leads in general to rational  and root--valued letters with real parameter
letters in the contributing alphabet, cf.~\cite{Ablinger:2017xml,ROOT1,ISR}.
In the case of the loop integrals one obtains letters of the kind
\begin{eqnarray}
\frac{1}{1-x(1-\eta)},~\frac{\sqrt{x(1-x)}}{\eta + x(1-\eta)},~\sqrt{x(1-\eta(1-x)},~~\eta \in [0,1].
\end{eqnarray}
The iterative integrals and constants which appeared in \cite{Ablinger:2017xml,Ablinger:2018brx}
could finally be all integrated to harmonic polylogarithms containing complicated arguments, at least up to 
one remaining integration, which allows their straightforward numerical evaluation.

In the case of phase space integrals with more scales, e.g. 
\cite{ISR},
also letters contribute, which may imply incomplete elliptic integrals and iterated structures thereof.
The integrands could 
not by rationalized completely by variable transformations, see also \cite{Besier:2018jen}. Contributing 
letters are e.g.
\begin{eqnarray}
\frac{x}{\sqrt{1-x^2}\sqrt{1-k^2x^2}},~
\frac{x}{\sqrt{1-x^2}\sqrt{1-k^2x^2}(k^2(1-x^2(1-z^2))-z^2)},
\end{eqnarray}
with $k, z \in [0,1]$. The corresponding iterative integrals are called Kummer--elliptic integrals. 
They are derived using the techniques described in Refs.~\cite{Ablinger:2014bra,RAAB1,GuoRegensburgerRosenkranz}.
\subsection{General properties of nested sums and iterated integrals}
\label{sec:43}

\vspace{1mm}\noindent
Iterated integrals obey shuffle relations
\begin{eqnarray}
\HA_{a_1,...a_m}(z) \cdot \HA_{b_1,...b_n}(z) &=&  \HA_{a_1,...a_n}(z) \shuffle \HA_{b_1,...b_n}(z) 
\nonumber\\ 
&=&
\sum_{c_i \in \{{a_1,...a_m} \shuffle {b_1,...b_n}\}} \HA_{c_i}(z).
\label{eq:shuf}
\end{eqnarray}
Here the order of the letter sequences of the quantities to be shuffled is preserved. The associated algebras
are called shuffle algebras \cite{REUTENAUER,Blumlein:2003gb}. The counting of the basis elements in the respective 
class \cite{RADFORD} may be done by counting its Lyndon words \cite{LYNDON} or using the  Witt-formulae \cite{WITT}.

Likewise, nested sums over hypergeometric terms form quasi--shuffle or stuffle \cite{Borwein:1999js} algebras 
\cite{HOFFMAN}. The stuffling relations are obtained by \cite{Moch:2001zr,Blumlein:2003gb}
\begin{eqnarray}
S_{a_1,...,a_n}(N) \cdot S_{b_1,...,b_m}(N) &=& 
\sum_{l_1=1}^N \frac{{\rm sign}(a_1)^{l_1}}{l_1^{|a_1|}}
S_{a_2,...,a_n}(l_1) \cdot S_{b_1,...,b_m}(l_1)
\nonumber\\ &+& 
\sum_{l_2=1}^N \frac{{\rm sign}(b_1)^{l_2}}{l_2^{|b_1|}}
S_{a_1,...,a_n}(l_2) \cdot S_{b_2,...,b_m}(l_2)
\nonumber\\ &-& 
\sum_{l=1}^N \frac{
({\rm sign}(a_1)
{\rm sign}(b_1))^{l_2}}{l^{|a_1|+|b_1|}}
S_{a_2,...,a_n}(l) \cdot S_{b_2,...,b_m}(l)
\nonumber\\
\end{eqnarray}
for harmonic sums and similar for the sums in extended spaces, 
see~\cite{Blumlein:2004bb,Blumlein:2009ta,Blumlein:2009fz,Ablinger:2013cf,Ablinger:2011te,Ablinger:2014bra}

These algebraic relations allow to reduced the number of contributing functions already significantly.

In the case of the iterated integrals different classes of mappings of the main argument may be used, which are 
helpful in many cases. The most important ones are
\begin{eqnarray}
k \cdot z \rightarrow z,~~k \in \mathbb{Q},~~~~1 - z \rightarrow z,~~~~\frac{1}{z} \rightarrow z,~~~~
\frac{1-z}{1+z} \rightarrow z.
\end{eqnarray}
Depending on the class of iterative integrals to be considered, not all of these relations map inside this class,
but can lead to functions in respective extensions. In these cases one just considers the wider space.
Iterated integrals also obey the differentiation relation
\begin{eqnarray}
\label{eq:DIFF}
\frac{d}{dz} \HA_{b,\vec{a}}(z) = f_b(z) \HA_{\vec{a}}(z).
\end{eqnarray}

Beyond the quasi--shuffle relations, also nested sums obey relations if considering their analytic continuation
to $N \in \mathbb{Q}, \mathbb{R}$ or $\mathbb{C}$, cf.~\cite{Blumlein:2004bb,Blumlein:2009ta,Blumlein:2009fz,
Ablinger:2013cf,Ablinger:2011te,Ablinger:2014bra}. These relations are called structural relations, 
cf.~\cite{Blumlein:2009ta}. 
The double- \cite{Blumlein:2009cf} and multiple arguments
relations and the differential relations, applied to the associated Mellin transforms, belong to this class.
The simplest double argument relation reads
\begin{eqnarray}
S_{n_1,...,n_p}(N) = 2^{n_1+n_2+...n_p-p} \sum_\pm S_{\pm n_1,...,\pm n_p}(2N)~,
\end{eqnarray}
for the harmonic sums.
Examples for the differential relation are
\begin{eqnarray}
\frac{d}{d N} S_{k}(N) &=& \frac{(-1)^{k-1}}{(k-1)!} \psi^{(k)}(N+1) = -k (S_{k+1}(N) - \zeta_{k+1})
\\
\frac{d}{d N} S_{-k}(N) &=& \frac{(-1)^{k-1}}{(k-1)!} \beta^{(k)}(N+1) = -k \left[S_{-(k+1)}(N) +\left(1 - 
\frac{\zeta_{k+1}}{2^k}
\right) \right],
\nonumber\\
\end{eqnarray}
with $\beta(N) = \left[\psi((N+1)/2) - \psi(N/2)\right]/2$. Therefore all single harmonic sums fall into a single
equivalence class under differentiation for $N$, which is represented by the harmonic sum $S_1(N)$.
The number of elements in the respective classes after applying the structural relations can also be counted 
by Witt--like formulae.

\subsection{Solutions in the case of non first order factorizable recurrences and differential operators}
\label{sec:44}

\vspace{1mm}
\noindent
Non--first order factorizing systems of differential or difference equations for the master integrals, cf. 
Section~\ref{sec:35}, occur at a certain order in massive Feynman diagram calculations. Well--known
examples for this are the sun--rise integral, 
cf.~e.g.~\cite{Broadhurst:1993mw,Laporta:2004rb,Bloch:2013tra,Adams:2013kgc,Adams:2014vja,Adams:2015gva,Adams:2015ydq},
the kite integral \cite{SABRY,Remiddi:2016gno,Adams:2016xah}, the three--loop QCD--corrections to the 
$\rho$--parameter 
\cite{Ablinger:2017bjx,Grigo:2012ji,Blumlein:2018aeq}, and the three--loop QCD corrections to the massive operator 
matrix element $A_{Qg}$ 
\cite{Ablinger:2017ptf}. In the case of the $\rho$--parameter a Heun equation \cite{HEUN} remains after separating 
the first order factorizing terms. Its solution can be given in terms of $_2F_1$--functions with a certain rational 
argument \cite{IVH,Ablinger:2017bjx} and rational parameters. These structures will later turn out not to occur 
accidentally. Next one may investigate  whether these solutions can be expressed in terms of complete elliptic 
integrals. This can be checked algorithmically using the triangle group \cite{TAKEUCHI}.

In the examples mentioned one can find representations in terms of complete elliptic integrals of the first 
and second kind, {\bf K} and {\bf E}, cf.~\cite{TRICOMI,WITWAT}. Here the question arises whether an argument 
transformation allows for a representation through only {\bf K}. It turns out that this in not possible in 
the present case according to the criteria given in \cite{Herfurtner1,Movasati1}. 

The homogeneous solution of the Heun equations are given by $_2F_1$--solutions $\psi_k^{(0)}(x), k=1,2$, at a 
specific rational argument. However, these integrals cannot be represented such that the variable $x$ just appears in
the boundaries of the integral. The inhomogeneous solution reads 
\begin{eqnarray}
\label{eq:INH}
\psi(x) = 
\psi^{(0)}_1(x)\left[C_1 - \int dx \psi_2^{(0)}(x) \frac{N(x)}{W(x)}\right] + \{1 \rightarrow 2\},
\end{eqnarray}
with $N(x)$ and $W(x)$ the inhomogeneity  and the Wronskian. $C_{1,2}$ are the integration constants. Through
partial integration the ratio $N(x)/W(x)$ can be transformed into an iterative integral. Since  
$\psi_{k}^{(0)}(x)$ cannot be written as iterative integrals, $\psi(x)$ is
obtained as an {\it iterative non--iterative integral} 
\cite{Blumlein:2016a,Ablinger:2017bjx} of the type
\begin{eqnarray}
\label{eq:Hit}
&& \hspace*{-8mm}
\mathbb{H}_{a_1,...,a_{m-1};{a_m,F_m(r(y_m))},a_{m+1},...a_q}(x) =
\nonumber\\ 	
&& \hspace*{-8mm}
\int_0^x dy_1 f_{a_1}(y_1) \int_0^{y_1} dy_2 ... \int_0^{y_{m-1}} dy_m f_{a_m}(y_m) F_m[r(y_m)] 
H_{a_{m+1},...,a_q}(y_m),
\end{eqnarray}
with $r(x)$ a rational function and $F_m$ a {\it non--iterative integral}. In general, usually more non--iterative 
integrals will appear in (\ref{eq:Hit}). $F_m$  denotes {\it any} non--iterative integral, implying 
a very general representation, cf.~\cite{Ablinger:2017bjx}.\footnote{This 
representation has been used in a more specific form also in \cite{Remiddi:2017har} later.}
In Ref.~\cite{Adams:2018yfj} an $\varepsilon$--form for the Feynman 
diagrams of elliptic cases has been found recently. Here transcendental letters contribute. This is in 
accordance with our earlier finding, Eq.~(\ref{eq:Hit}), which, as well is an iterative integral over all objects 
between the individual iterations and to which now also the non--iterative higher transcendental functions 
$F_m[r(y_m)]$ contribute. One may obtain fast convergent representations of $\mathbb{H}(x)$ by overlapping series 
expansions around $x = x_0$ outside possible singularities, see Ref.~\cite{Ablinger:2017bjx} for details.

Now we return to the elliptic case. Here one one may transform the kinematic variable $x$ occurring as
${\rm \bf K}(k^2) = {\rm \bf K}(r(x))$ into the variable $q = \exp[i\pi \tau]$ analytically with
\begin{eqnarray}
\label{eq:EL1}
k^2 = r(x) = \frac{\vartheta_2^4(q)}{\vartheta_3^4(q)},
\end{eqnarray}
by applying a cubic order Legendre--Jacobi transformation, where $\vartheta_l, l=1,...,4$ denote Jacobi's 
$\vartheta$-functions and ${\sf Im}(\tau) > 0$.  In this way Eq.~(\ref{eq:INH}) 
is rewritten in terms of the new variable. The integrands are given by products of meromorphic modular 
forms, cf.~\cite{SERRE,COHST,ONO1}, which can be written as a linear combination of ratios of Dedekind's 
$\eta$-function
\begin{eqnarray}
\label{eq:EL2}
\eta(\tau) = q^{\tfrac{1}{12}} \prod_{k=1}^\infty (1-q^{2k})~.
\end{eqnarray}
Depending on the largest multiplier $k \in \mathbb{N}$, $k_m$, of $\tau$ in the argument of the $\eta$-function, 
the solution transforms under the congruence subgroup $\Gamma_0(k_m)$ and one can perform Fourier expansions in $q$ 
around the different cusps of the problem, cf.~\cite{ZUDILIN,BROADH18}. 

For holomorphic modular forms, one obtains representations in Eisenstein 
series with character, while in the meromorphic case additional $\eta$--factors in the denominators are present.
In the former case the $q$--integrands can be written in terms of elliptic polylogarithms 
in the representation \cite{Adams:2014vja,Adams:2015gva}
\begin{eqnarray}
\label{eq:EL3}
{\rm ELi}_{n,m}(x,y) = 
\sum_{k=1}^\infty
\sum_{l=1}^\infty \frac{x^k}{k^n} \frac{y^l}{l^m} q^{k l}
\end{eqnarray}
and products thereof, cf.~\cite{Adams:2015gva}. The corresponding $q$--integrals can be directly performed.
The solution (\ref{eq:INH}) usually appears for single master integrals. Other master integrals are obtained
integrating further other letters, so that finally representations by $\mathbb{H}(x)$ occur.
Iterated modular forms, resp. Eisenstein series, have been also discussed recently in 
\cite{Adams:2017ejb,Broedel:2018iwv,ELL1}.

Returning to the example of the $\rho$--parameter we find that it cannot be represented in terms of elliptic 
polylogarithms only because of the emergence of the complete elliptic integral {\bf E}, for which the singularity 
in $q$ implies Dedekind $\eta$--functions appearing in the denominator. These factors have no (known) closed form 
$q$--expansion, cf.~\cite{RADEM}. Let us also remark that the corresponding non--iterative solutions are sometimes 
found mapping first into the non--physical region. In the end one has to perform  an analytic continuation back to 
the physical case, which requires to have closed form expressions.
Recent developments in the field of Feynman integrals and elliptic structures are discussed in 
Refs.~\cite{WEINZIERL,BROEDEL}.

Let us mention that in some applications also non--factorizable differential equations of 3rd order and higher
can occur. The higher the order the less is known about the analytic structure of the solutions in the general case.
In the future one will be confronted with these cases and practical solutions for them have to be found, including
highly precise numerical representations in the physical cases. This issue is presently under study.
\subsection{Spaces of special numbers}
\label{sec:45}

\vspace{1mm}\noindent
For the sums of Section~\ref{sec:41} which are convergent in the limit $N \rightarrow \infty$ and the 
iterated integrals of Section~\ref{sec:42} which can be evaluated at $x=1$ one obtains two sets of special 
numbers. They span the solution spaces for zero--scale quantities and appear as boundary values 
for single--scale problems. Examples for these special numbers are the multiple zeta values 
\cite{Blumlein:2009cf}, associated to the harmonic sums and harmonic polylogarithms, special generalized 
numbers \cite{Ablinger:2013cf} like $\Li_2(1/3)$ and  $\Li_k(-1/2)$, cf.~\cite{SQALPHA}, 
associated to generalized sums and to Kummer--Poincar\'e iterated 
integrals, special cyclotomic numbers \cite{Ablinger:2011te} like Catalan's number, special binomial numbers 
\cite{Ablinger:2014bra}, as e.g. arccot($\sqrt{7}$),
and special constants in the elliptic case \cite{Ablinger:2017bjx,Laporta:2017okg}. The latter numbers are given by 
integrals
involving complete elliptic integrals at special rational arguments and related functions.
In general these numbers obey more relations than the finite sums and iterated integrals. One may use the 
PSLQ--method to get a first information on relations between these numbers occurring in a given problem and proof
the conjectured relations afterwards.

\subsection{Numerical representations}
\label{sec:46}
\vspace{1mm}
\noindent
Physical observables based on single scale quantities can either be represented in Mellin $N$--space
or $x$--space. Many of these representations are given in terms of either nested sums or iterative integrals.
However, there are also contributions due to iterative non--iterative integrals.

Representations in Mellin $N$--space allow the exact analytic solution of evolution 
equations \cite{Blumlein:1997em} and scheme-invariant evolution equations can be derived in this way 
\cite{Blumlein:2000wh,Blumlein:2004xs}. The $x$--space representation is then obtained by a single 
numerical integral around the singularities of the respective quantity for $N  \in \mathbb{C}$, cf. 
\cite{Blumlein:1997em}, requiring to know the complex representation of the integrand in $N$--space.

In the case of harmonic sums semi--numerical representations were given in \cite{Blumlein:2000hw,Blumlein:2005jg}. 
Furthermore, it is known that the basic harmonic sums, except of $S_1(N)$, which is represented by the Digamma 
function, and its polynomials, have a representation by factorial series \cite{FACT1,FACT2}, which has been used 
in \cite{Blumlein:2009fz,Blumlein:2009ta} for their asymptotic representation, see also \cite{Kotikov:2005gr}.

The asymptotic representation of these quantities is thus given to arbitrary precision and one may use
the recurrence relations of these quantities to analytically continue (\ref{eq:SUM}) from integer values of $N$ to 
$N \in \mathbb{C}$. Here it is important to observe the crossing relations for the respective process 
\cite{Politzer:1974fr,Blumlein:1996vs}
which either implies the analytic continuation from the even or from the odd integers.
These steps also apply to the other types of sums which were described in 
Refs.~\cite{Ablinger:2011te,Ablinger:2013cf,Ablinger:2014bra,SQALPHA} analogously, which appear in certain physical 
problems, cf.~\cite{Ablinger:2014yaa,Ablinger:2015tua}. 

In the case that the corresponding relations are not given in tabulated form, they can be calculated using the 
package {\tt HarmonicSums}.
Relations for harmonic sums are also implemented in {\tt summer} 
\cite{Vermaseren:1998uu}, and for generalized harmonic sums in {\tt nestedsums} \cite{Weinzierl:2002hv}, {\tt 
Xsummer} \cite{Moch:2005uc}, and {\tt PolyLogTools} \cite{Duhr:2019tlz}.

In other applications one may want to work in $x$--space directly. Here numerical representations are available 
for the Nielsen integrals \cite{NIELSEN}, the harmonic polylogarithms 
\cite{Gehrmann:2001pz,
Vollinga:2004sn,
Maitre:2005uu,
Ablinger:2018sat}, the Kummer--Poincar\'e 
iterative integrals 
\cite{Vollinga:2004sn}, the cyclotomic harmonic polylogarithms \cite{Ablinger:2018zwz}, and those implied by 
quadratic forms \cite{SQALPHA}. These representations are 
also useful to lower the number of numerical integrations for more general problems, e.g. in the multivariate case.
The relations for the corresponding quantities are implemented for the harmonic polylogarithms in 
\cite{Remiddi:1999ew,Maitre:2005uu} and for all iterative integrals mentioned, including general iterative integrals, in
the package {\tt HarmonicSums}. Moreover, the packages {\tt summer} \cite{Vermaseren:1998uu}, the multiple zeta 
values data mine \cite{Blumlein:2009cf}, and {\tt HarmonicSums} also provide extensive lists of special numbers in 
various tabulated basis representations allowing quick numerical evaluation. Dynamical numerical evaluations are 
provided by the package described in \cite{Vollinga:2004sn} and by {\tt HarmonicSums} for  non tabulated cases.

Finally, we remark that numerical evaluations of a series of elliptic integral solutions were given in 
Refs.~\cite{ELNUM,Bogner:2017vim}.
\section{Precision goals in testing the Standard Model}
\label{sec:5}

\vspace{1mm}\noindent
I would like to finally discuss the application of the mathematical methods described to precision
prediction for dedicated observables in Quantum Field Theory, which are measured at high precision
both in low energy experiments and at present and future colliders.

At low energies  central quantities are $(g-2)_{e,\mu}$ \cite{GM2}, for which the  $O(\alpha^5)$ 
contributions have been computed numerically \cite{Aoyama:2012wj} and the 
$O(\alpha^4)$ contributions and parts of $O(\alpha^5)$ terms have been calculated analytically 
\cite{Baikov:2013ula,Kurz:2015fhj,Kurz:2016bau,Marquard:2017iib,Laporta:2017okg,Volkov:2019phy}.
For various years there is a discrepancy between the experimental data and the theoretical prediction.
For massive calculations also the on--shell renormalization and decoupling constants are 
important. At present highest loop order they were given in Refs.~\cite{OSR}.  

In massless QCD the level of 3--loop corrections for the anomalous dimensions and Wilson coefficients
has been reached \cite{ANOMDWIL,Vermaseren:2005qc,Ablinger:2017ptf,Behring:2019tus}. The corrections to the 
$\beta$-function \cite{BETA} are available to at five loop order. 
The heavy flavor corrections to deep--inelastic 
structure functions reached the level of 3--loop corrections 
\cite{Bierenbaum:2009mv,HQDIS,Ablinger:2018brx,Ablinger:2017xml,Ablinger:2017ptf} and 
are on the way to be 
completed. Furthermore, there are also analytic 3-loop corrections to the inclusive Higgs production 
rate \cite{Anastasiou:2015vya} and the Drell-Yan process \cite{Duhr:2020seh,Duhr:2020sdp}, while the 
NNLO corrections for the $t\bar{t}$-production in hadronic collisions \cite{Czakon:2013goa} has been 
computed numerically, because of the presence of more involved integrals, still to be solved analytically.

All these processes are essential to pin down the accuracy of the parton distribution functions
in the region of a clear twist--2 dominance \cite{PDF}, also accounting for jet production 
cross sections in $pp \rightarrow Z + \text{jet}$ at NNLO \cite{Ridder:2015dxa,Boughezal:2015ded} and 
in $ep$ two--jet production \cite{EPJET}.

The final goal is here the precision measurement of the strong coupling constant $\alpha_s(M_Z^2)$ in a
widely unique manner. This can also be achieved using the method of scheme-invariant evolution equations
\cite{SCHINV} for which the initial conditions are measured directly. Furthermore, one would like to 
determine at least the charm quark mass, $m_c$ \cite{Alekhin:2012vu}, in a correlated way with the parton densities 
and 
$\alpha_s(M_Z^2)$. Here one wants to reach relative accuracies of the order of 1\% and better.
It finally may be necessary to study QCD evolution at the level of N$^3$LO \cite{Moch:2017uml}, in 
particular if one wants to include small $x$ effects and check the analytic predictions of the BFKL formalism 
\cite{BFKL}.

Facing future colliders such as the FCC$\_$ee \cite{Abada:2019zxq} a measurement of the fine structure 
constant $\alpha(M_Z^2)$ is possible at very high accuracy by using the forward--backward asymmetry 
\cite{Janot:2015gjr} and one needs to know precision predictions on the QED radiative corrections 
\cite{Blumlein:2021jdl}. This measurement may
yield an independent access to the size of the hadronic contributions to $\alpha$. In studying the
$Z$ resonance at the FCC$\_$ee one expects precisions of $\sim$~100~keV for $M_Z$ and the 
width of the $Z$ boson, 
$\Gamma_Z$, which requires refined QED corrections. Those for the initial state radiation have been 
calculated in \cite{ISR} using a wide host of methods described in this article. There are more 
goals, as e.g. the precision understanding of the top-threshold, cf. e.g. 
\cite{Beneke:2016kkb,Beneke:2017rdn,Beneke:2015kwa,Bach:2017ggt,Hoang:2013uda,
Seidel:2013sqa,Simon:2016pwp}, the measurement of 
${\rm sin}^2 \theta_W$ \cite{Dubovyk:2019szj}, and the precision measurement of the $W$-boson mass. 

Beyond the more inclusive measurements we have described, there is a large list of hard exclusive 
reactions needed in the analysis of the experimental data at the LHC and at future colliders.
These corrections require a lot more numerical technologies, because of the number of different scales 
present. For a recent summary of the status see \cite{Heinrich:2020ybq}. 

Effective field methods can 
also be applied to classical gravity to derive higher order post--Newtonian corrections for the inspiraling
process of two massive objects. With these methods currently the level of the 5th post--Newtonian order 
has been reached \cite{PN5}.
\section{Conclusions}
\label{sec:6}

\vspace{1mm}\noindent
With the progress in analytic precision calculations in Quantum Field Theory more and more 
mathematical technologies are used to solve the corresponding integrals analytically. The 
classical and Nielsen polylogarithms turned out to be not sufficient any more to represent
intermediary and the final results in the late 1990ies. Moreover, the method of hypergeometric
functions, which has fully provided the corresponding integral representations up this point
failed to cover more involved structures. This applied already to massless and massive
calculations for single scale  quantities in QCD at 3--loop order. The first indication for 
this was that the arguments appearing in the 2--loop Nielsen integral representations became
more and more complicated. At this time  it has also been discovered that  
Mellin-space representations lead to essential compactifications \cite{Vermaseren:1998uu,Blumlein:1998if}
and later it turned out that all the single--scale 2--loop result can be written in terms of just 
six harmonic sums \cite{Blumlein:2004bb,Blumlein:2005im,Blumlein:2006rr}.

The iterative integral structure has been known from the classical polylogarithms \cite{POLYLOG}
and Nielsen integrals \cite{NIELSEN} and led to the harmonic polylogarithms \cite{Remiddi:1999ew}.
During the following years more and more of these structures have been revealed. Here the difference ring techniques
\cite{DRING,Schneider:2007a,Schneider:2013a,SCHNEIDER} played an essential role, since the corresponding
structures were found constructively, mostly in massive calculations. The Mellin transform of these
quantities allowed then to find the associated iterated integrals.

For some years now also iterative non--iterative integrals are known and were widely studied in the case
of complete elliptic integrals. However, we expect more involved structures to emerge. Possible new
structures of this kind could be Abel--integrals \cite{NEUMANN} and integrals  related to 
K3--surfaces \cite{BSCH}. 

In massive calculations \cite{Ablinger:2017ptf,Blumlein:2019oas} we observe a growing number of
non first order factorizable recurrences, probably already containing structures beyond the elliptic
level. Their solution calls for a general method, which might provide semi--analytic numerical 
representations in the end, which can be tuned to any precision.
Yet one is also interested in the concrete mathematical structures of these cases. Global methods like
the recurrences or the method of differential equations will have a hard time to reveal those.
It is rather important here to analyze the multidimensional integrands first, which is provided by
applying cutting techniques in a systematic manner, performing various Hilbert-transforms 
\cite{Hilbert:1912,KRONIG,KRAMERS}. This has been successfully practiced at one--loop order, 
see e.g.~\cite{Abreu:2017mtm}, and also revealed in a nice manner the emergence of elliptic 
integrals.\footnote{To see the same on the basis of a Heun differential equation is much more difficult 
\cite{Ablinger:2017bjx}.} 
The method has been advocated early by M.~Veltman in his PhD thesis
\cite{Veltman:1963th}, see also \cite{REMI1}.

In the future we will see an intense cooperation of theoretical physicists, mathematicians and scientists
working on large scale computer algebra on the topic of the calculation of Feynman integrals by
antidifferentiation. The field will conquer new horizons, one never thought of. All the participating 
fields will enormously profit from this work and new masterpieces of the esprit humain will be seen.

\vspace{2ex}
\noindent
{\bf Acknowledgment.}~I would like to thank to all who have contributed to the present volume 
and all my colleagues with whom I have had countless fruitful discussions on the present topic 
during the last 30 years. This project has received funding from the European Union's Horizon 
2020 research and innovation programme under the Marie Sk\l{}odowska--Curie grant agreement 
No.~764850, SAGEX.

\endgroup
\end{document}